# Propagation dynamics of abruptly autofocusing circular Airy-Gaussian vortex beams in the fractional Schrödinger equation

Shangling He[1], Boris A. Malomed[2,3], Dumitru Mihalache[4], Xi Peng[5], Xing Yu[1], Yingji He[5#], and Dongmei Deng[1]*

[1]*Guangdong Provincial Key Laboratory of Nanophotonic Functional Materials and Devices, South China Normal University, Guangzhou 510631, China*

[2] *Department of Physical Electronics, School of Electrical Engineering, Faculty of Engineering, and Center for Light-Matter Interaction, Tel Aviv University, Tel Aviv 69978, Israel*

[3] *Instituto de Alta Investigación, Universidad de Tarapacá, Casilla 7D, Arica, Chile*

[4]*Horia Hulubei National Institute for Physics and Nuclear Engineering, P.O. Box MG-6, RO-077125, Bucharest-Magurele, Romania*

[5] *School of Photoelectric Engineering, Guangdong Polytechnic Normal University, Guangzhou 510665, China*

Emails: #heyingji8@126.com, and *dmdeng@263.net (corresponding author)

Abstract

We introduce axisymmetric Airy-Gaussian vortex beams in a model of an optical system based on the (2+1)-dimensional fractional Schrödinger equation (FSE), characterized by its Lévy index (LI), $1 < \alpha \leq 2$. By means of numerical methods, we explore propagation dynamics of the beams with vorticities from 0 to 4. The propagation leads to abrupt autofocusing, followed by its reversal (rebound from the center). It is shown that LI, the relative width of the Airy and Gaussian factors, and the vorticity determine properties of the autofocusing dynamics, including the focusing distance, radius of the focal light spot, and peak intensity at the focus. A maximum of the peak intensity is attained at intermediate values of LI, close to $\alpha = 1.4$. Dynamics of the abrupt autofocusing of Airy-Gaussian beams carrying vortex pairs (split double vortices) is considered too.

## 1. Introduction

Abruptly autofocusing optical beams are subjects of broad interest in theoretical [1] and experimental [2,3,4] studies. Being able to sharply concentrate their energy at focal points, such beams offer potential applications to biomedical therapy [1], generation of "light bullets" [5], design of precise optical tweezers [3,6], stimulating photo-polymerization [7], etc. Similar phenomena were also predicted for matter waves in Bose-Einstein condensates [8].

In this context, autofocusing ring Airy beams have drawn broad interest due to their ability to abruptly increase the intensity at the focal point by several orders of magnitude [9-20]. Roughly speaking, in this case the autofocusing is driven by the effective surface tension of the narrow annular area that carries the optical power, see Figs. 1(b1), 4(a1,a2), and 7(b1) below. This property may be used, in particular, to focus the beam on a selected target in biomedical applications, keeping low intensity around the focus, so as to avoid damaging surrounding tissues. Properties of the abrupt autofocusing, such as the focal intensity and position, followed by its reversal (defocusing), may be controlled by parameters of the setting, one of which is the "distribution factor" *b*, which determines a ratio between widths of the Airy and Gaussian factors in the input, see Eq. (2) below.

The above-mentioned research was performed in the framework of the standard Schrödinger equations. More recently, based on the path-integral approach, a fractional generalization of the Schrödinger equation, developed in the framework of quantum and statistical mechanics, was proposed by Laskin [21]. The so derived fractional Schrödinger equations (FSEs) have drawn much interest in various areas of physics and applied mathematics [22-31]. In particular, they govern the evolution of fields in fractional-dimension spaces, and dynamics of particles with fractional spin [32]. Due to issues with handling nonlocal operators that represent fractional derivatives and Laplacians, characterized by the respective Lévy index (LI) [33], and scarcity of relevant experimental results, the advancement in this area was rather slow. An essential step forward was made by Longhi [34], who has introduced FSE into optics and obtained solutions for dual Airy beams with off-axis longitudinal pumping in spherical optical cavities. The realization of the FSE theory in optical fields provides abundant possibilities for studies of the fractional-order beam-propagation dynamics. Subsequently, the propagation of beams was investigated in FSE with various external potentials and nonlinear terms and in other fractional evolution equations [35-44].

In particular, the propagation of ring Airy beams modeled by FSE has been explored numerically [45]. However, to the best of our knowledge, the propagation of autofocusing circular Airy-Gaussian vortex (AGV) beams in FSE optical system, which is an issue of straightforward physical interest, has not been addressed in previous works, which is the subject of the present work.

The paper is organized as follows. In Section 2, we introduce the model for the AGV beams in the FSE optical system. The propagation of abruptly autofocusing AGV beams in the FSE optical system, which is followed by defocusing, is systematically studied in Section 3. The analysis is performed for the AGVs with all values of the intrinsic vorticity (winding number) from 0 to 4 (i.e., both zero- and higher-vorticity beams are included). Further, autofocusing and defocusing of a pair of AGVs, which may be considered as a double vortex in a split state, is addressed in Section 4. The paper is concluded by Section 5.

## 2. The model

The propagation of AGV waves in a linear medium is governed by FSE, written in the following form [45]:

$$i\frac{\partial u}{\partial z} - \frac{1}{2kw_0^{2-\alpha}}\left(-\frac{\partial^2}{\partial x^2} - \frac{\partial^2}{\partial y^2}\right)^{\alpha/2} u = 0, \qquad (1)$$

where $u$ is the amplitude of the optical wave, $x$ and $y$ are the scaled transverse coordinates, $z$ is the propagation distance, and $\alpha$ is the LI ( $1 < \alpha \leq 2$ ). The fractional-diffraction operator in Eq. (1) is defined by the known integral expression [21-23],

$$\frac{1}{2w_0^{2-\alpha}}\left(-\frac{\partial^2}{\partial x^2} - \frac{\partial^2}{\partial y^2}\right)^{\alpha/2} u(x,y,z)$$
$$\equiv \frac{1}{8\pi^2 w_0^{2-\alpha}} \iint dk_x dk_y \, \exp(ik_x x + ik_y y)\left(k_x^2 + k_y^2\right)^{\alpha/2} \hat{u}(k_x, k_y, z),$$

where

$$\hat{u}(k_x, k_y, z) = \iint dx dy \, \exp(-ik_x x - ik_y y) u(x,y,z)$$

is the two-dimensional Fourier transform of the field, which is a function of wave numbers $k_x$ and $k_y$. With $\alpha = 2$, Eq. (1) amounts to the standard two-dimensional linear Schrödinger equation.

We aim to solve Eq. (1) with the input in the form of the AGV beam written in terms of the polar coordinates, $(r, \phi)$:

$$u(r, \phi, z=0) = A_0 \text{Ai}\left(\frac{r_0 - r}{bw}\right) \exp\left(d \frac{r_0 - r}{bw}\right) \exp\left[-\frac{(r_0 - r)^2}{w^2}\right]\left(\frac{r^m}{w^m} e^{im\phi}\right), \qquad (2)$$

where $A_0$ is its amplitude, Ai is the standard Airy function, $r_0$ determines the radius of the primary Airy ring, $w$ is the width of the Gaussian factor, the parameter $b$ is the "distribution factor", i.e., the ratio of widths of the Airy and Gaussian factors, which determines what factor dominates in the AGV wave form, $m = 0, 1, 2, ...$ is the integer vorticity of the optical vortex, and $0 \leq d < 1$ is the exponential truncation factor of the Airy wave.

Equation (1) with initial condition (2) was solved numerically, by means of the fast-Fourier-transform method. In the simulations, we set $A_0 = 1$ and $d = 0.2$ (this value implies that the Airy function is not truncated too much, being able to feature its essential structure). Other coefficients correspond, in terms of the realization in optics, to the carrier wavelength $\lambda = 532 \times 10^{-6}$ mm and waists $r_0 = w = 1$ mm, while LI $\alpha$ and distribution factor $b$ remain free parameters. These values of $\alpha$ and $r_0 = w$ determine the characteristic propagation scale, i.e., the Rayleigh distance, $Z_R = kw^2 / 2 \approx 6$ m.

## 3. Numerical results

### 3.1. Self-focusing of the AGV beams and its reversal

We first address the propagation of abruptly autofocusing AGV beams in the FSE optical model with values of free parameters that adequately represent a generic situation: $\alpha = 1.4$ and $b = 0.1$, and we focus on unitary vortices, with $m = 1$ (values $m \geq 2$ are considered below in Figs. 5-7). For this case, the results of the simulations are displayed in Fig. 1. The axisymmetric field distribution demonstrates accelerated autofocusing to the smallest focal spot, which is attained at $z = 1.2Z_R$, which is followed by the reversal, i.e., rebound from the focus. The reversal qualitatively resembles, in particular, a similar effect produced by usual one-dimensional Schrödinger equation with the third-order dispersion and input in the form of the truncated Airy function [46].

The autofocusing being driven by the effective surface tension (alias elasticity) of the narrow ring-shaped area, it should be arrested and switch to the rebound when the ring's thickness becomes comparable to its overall radius. This expectation is corroborated by the picture displayed in Fig. 1(a), see also similar effects in Figs. 2(a,b) and Figs. 3(a,b) below. In addition to that, the positive contribution of vorticity $m$ to the FSE's Hamiltonian contributes to the arrest of autofocusing, therefore the value of the radius at the rebound point increases with the growth of $m$, as is seen below in Fig. 5(c) and Figs. 6(a)-6(e).

The wave profile of the present solution at the expansion stage resembles the Bessel-beam shape. In the course of the evolution, the beam keeps the vortical structure of the phase field rotating in the counter-clockwise direction (i.e., the scroll structure, in the three-dimensional rendition), and the respective hole in the amplitude distribution.

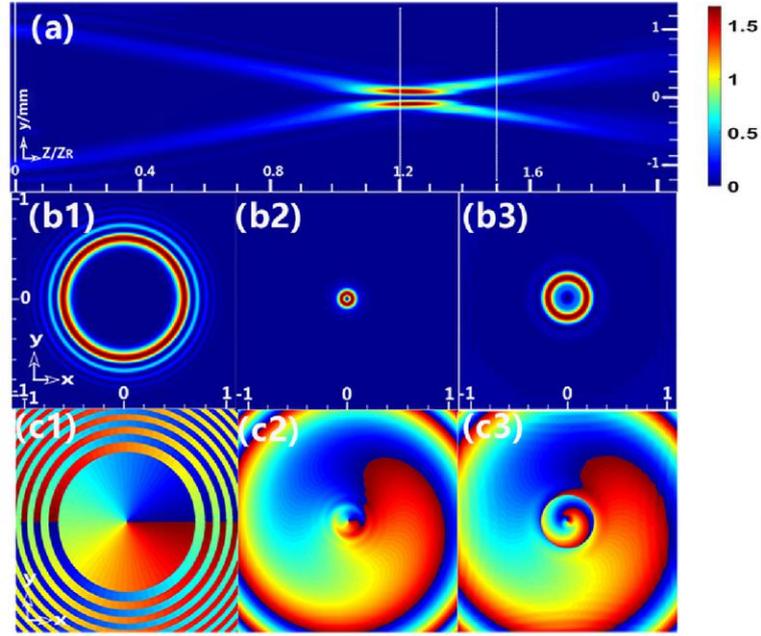

Fig. 1. (color online) (a) A cross section, in the plane of $y=0$, of the numerically simulated propagation of the AGV beam, produced by input (2) with parameters $\alpha=1.4, b=0.1, m=1$ in Eqs. (1) and (2), displayed by means of the heatmap of $|u(x,y=0,z)|^2$. Panels (b1)-(b3) are snapshots of the transverse intensity patterns, $|u(x,y,z)|^2$, at values of $z$ marked by dashed vertical lines in (a). (c1)-(c3) Snapshots of the phase distributions corresponding to panels (b1)- (b3).

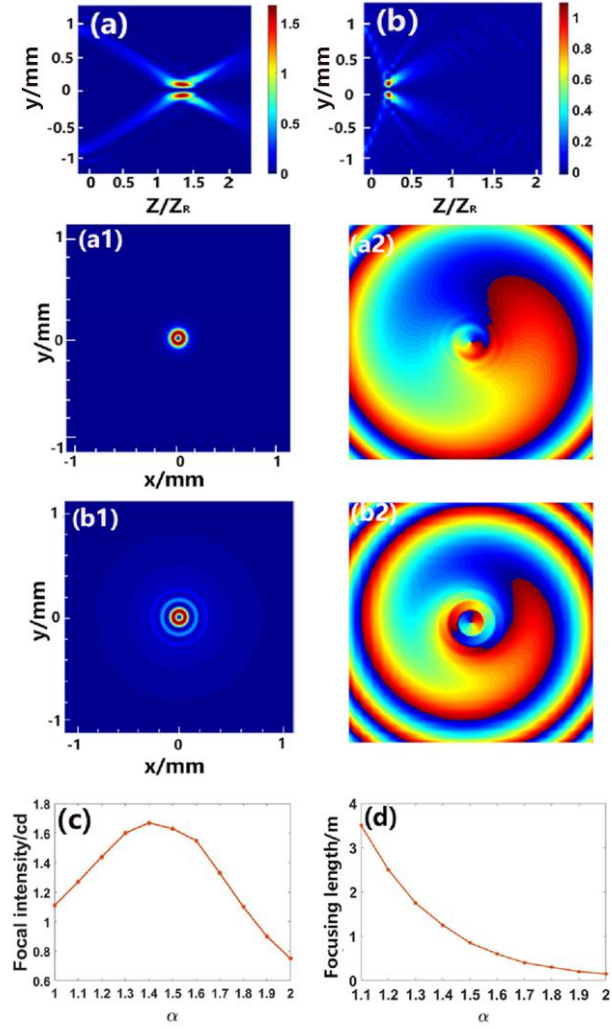

Fig. 2. (color online) Side-view propagation of the Airy-Gaussian vortex beams in NLSE optical system under the LI (Lévy index): (a) $\alpha = 1.4$; (b) $\alpha = 1.8$. Panels (a1,a2) and (b1,b2): the same as in Figs. 1(b2,c2), for $\alpha = 1.4$ and $\alpha = 1.8$, respectively. Panels (c) and (d): the focal intensity and focusing length vs. LI. Other parameters are $m = 1$ and $b = 0.1$. Panels (a), (a1), and (a2) are actually tantamount to ones (a), (b2), and (c2), respectively, in Fig. 1. These panels are included here for the sake of comparison with the results obtained for $\alpha = 1.8$.

### 3.2. Control of the autofocusing by the Lévy index (LI).

The effect of the variation of LI on the autofocusing dynamics is presented in Fig. 2. Comparison of panels 2(a) and 2(b) demonstrates that larger LI leads to faster autofocusing, which is natural, as the autofocusing is driven by the diffraction term in Eq. (1), which becomes stronger with the increase of $\alpha$ [the case of $\alpha = 1.8$, shown in Fig. 2(b), is nearly tantamount to the usual Schrödinger equation, which corresponds to $\alpha = 2$ ]. Further, for the same reason the ring pattern in panel 2(a1), corresponding to $\alpha = 1.4,$ is tighter than its counterpart in 2(b1), which is plotted for $\alpha = 1.8.$

In a systematic form, the dependence of characteristics of the autofocusing dynamics on LI are shown in Figs. 2(c) and 2(d). The monotonous decrease of the focusing length with the increase of $\alpha$ in Fig. 2(d) has the same natural explanation as mentioned above, i.e., enhancement of the diffraction term that drives the autofocusing. On the other hand, a nontrivial feature observed in Fig. 2(c), which does not find a simple explanation, is that the peak intensity at the focal spot attains a maximum at a particular value of LI, *viz.*, $\alpha \approx 1.4,$ and decreases with the further increase of LI.

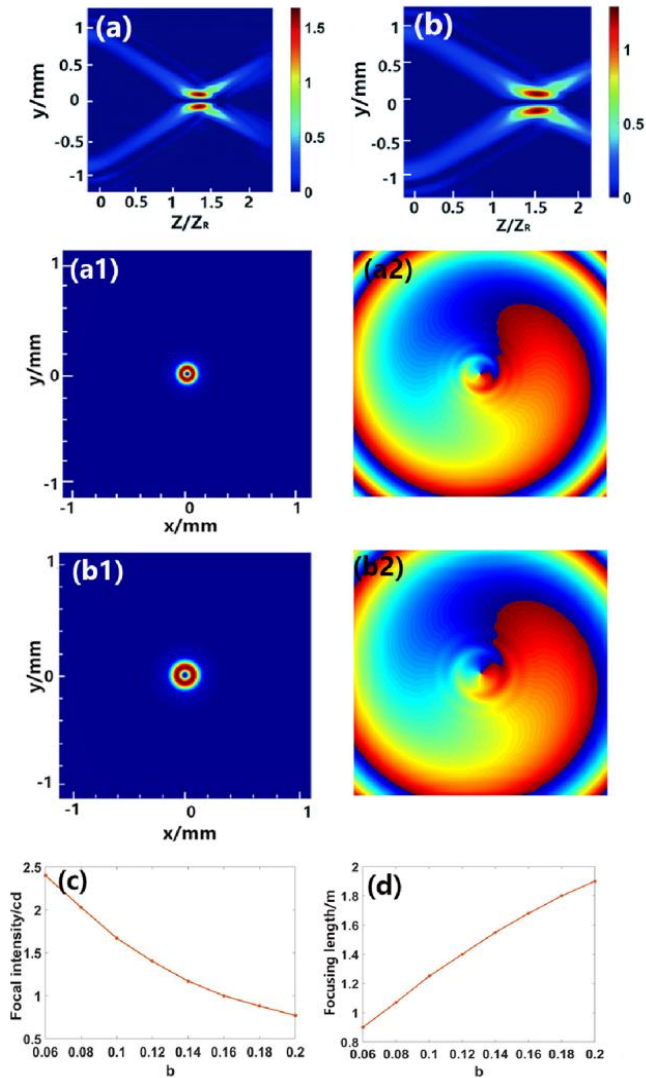

Fig. 3. (color online) Side-view propagation of the Airy-Gaussian vortex beams, but for different values of the distribution factor: $b = 0.1$ in panels (a,a1,a2), and $b = 0.13$ in (b,b1,b2). Panels (c) and (d) show the peak intensity at the focal spot and focusing length as functions of $b$. Other

parameters are $m=1$ and $\alpha=1.4$.

### 3.3. Control of autofocusing by the distribution factor (the width ratio of the Airy and Gaussian factors)

Figure 3 presents the effect of the variation of the distribution factor, i.e., the parameter $b$ in Eq. (2), on the autofocusing dynamics. It is seen that the effect is quite strong even if $b$ varies keeping relatively small values [hence the Airy factor remains a dominant one in input (2)]. In particular, the tightest autofocusing is attained at propagation distances $z=1.25Z_R$ and $z=1.5Z_R$ for $b=0.1$ and $b=0.13$, respectively. The rapid decrease of the peak intensity in Fig. 3(c) and increase of the focusing distance in 3(d) with the increase of $b$ are explained by the fact that larger $b$ enhances the attenuation of the input signal (2) by making the Gaussian factor effectively stronger.

The results displayed in Fig. 3 for $b=0.1$ and $b=0.13$ at fixed values of the LI and vorticity, $\alpha=1.4$ and $m=1$, are further illustrated in Fig. 4, by displaying the intensity profiles of the input and those at the tightest-focusing point. For both values of $b$, the peak intensity at the latter point exceeds the initial value by a factor $\approx 11$.

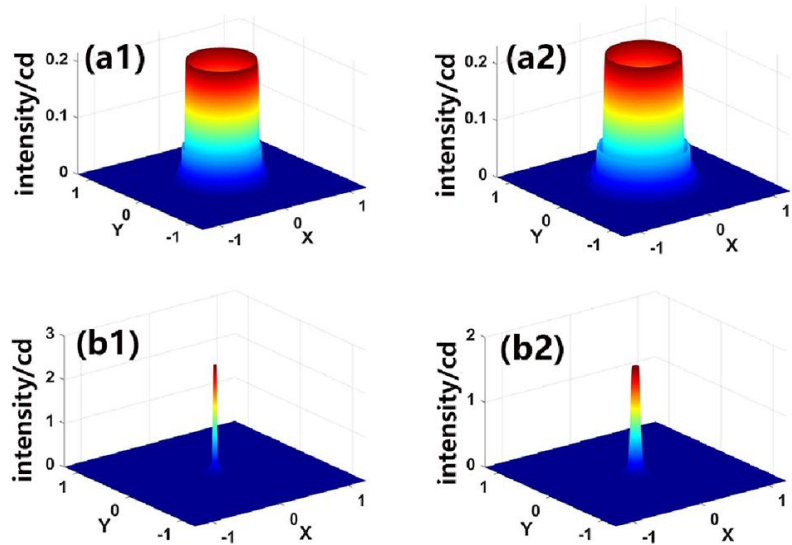

Fig. 4. (color online) Intensity distributions in the input and in the tightest autofocused state for $b=0.06$ (a1,b1) and $b=0.1$ (a2,b2). Other parameters are $m=1, \alpha=1.4$. Note the difference in the scales of the vertical axes between the top and bottom panels.

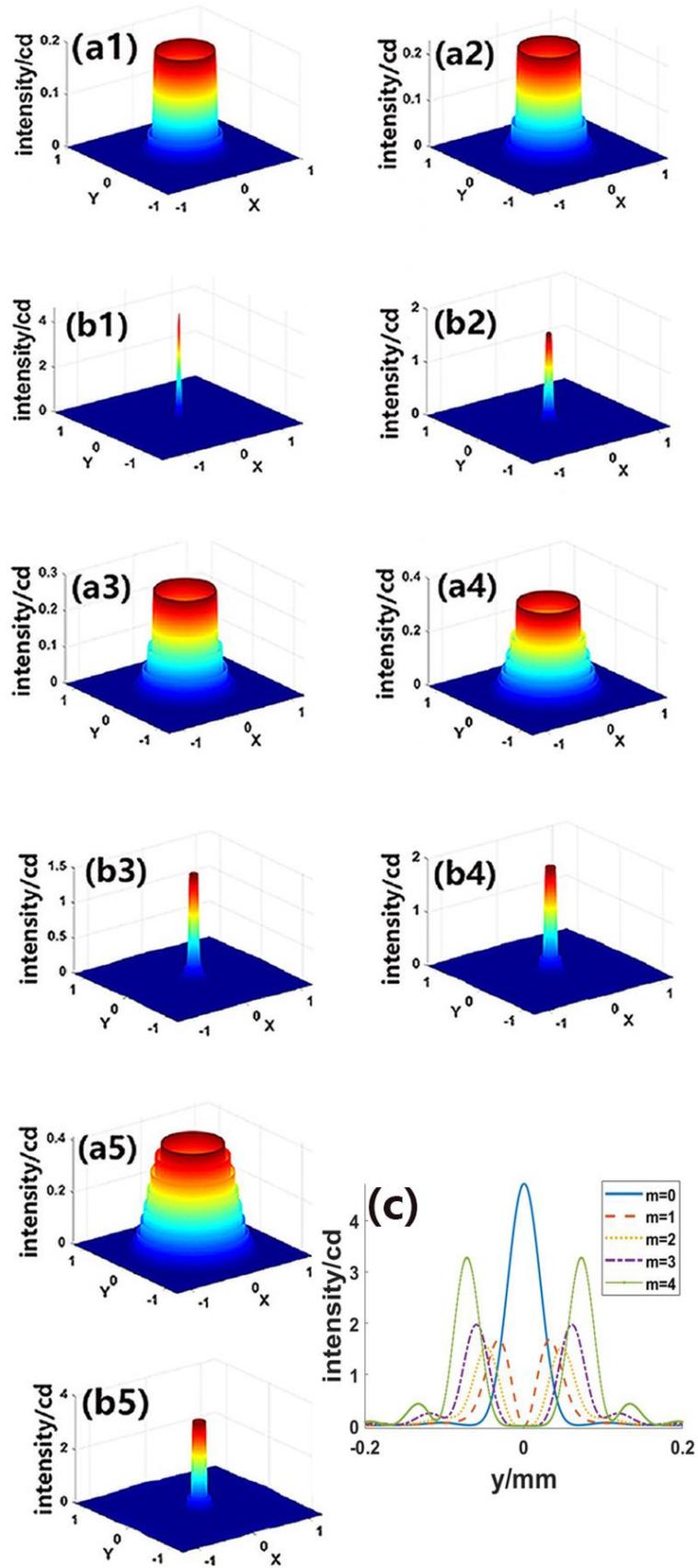

Fig. 5. (color online) Panels (a1-a4) and (b1-b4) display the same as in Fig. 4, but for different vorticities (winding numbers) of the input in Eq. (2): (a1) and (b1) $m = 0,$ (a2) and (b2) $m = 1,$

(a3) and (b3) $m = 2$, (a4) and (b4) $m = 3$, (a5) and (b5) $m = 4$. (c) A juxtaposition of cross-sections of these intensity profiles in the tightest autofocused states. The other parameters are $\alpha = 1.4$ and $b = 0.1$.

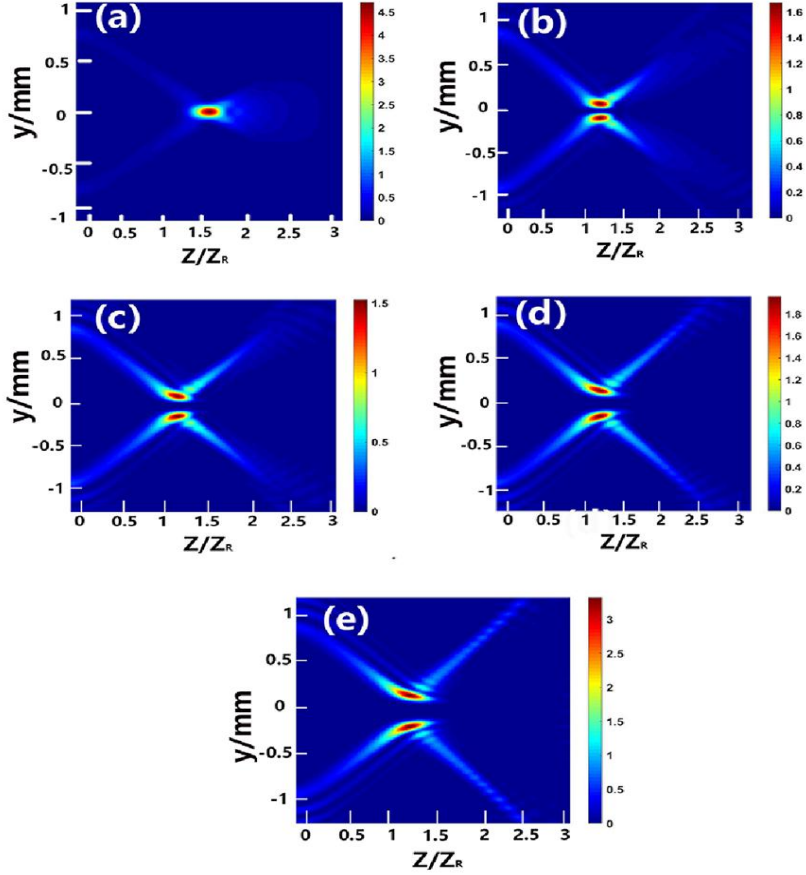

Fig. 6. (color online) Cross sections showing the autofocusing-defocusing dynamics of the same AGV beams as in Fig. 5: (a) $m = 0$, (b) $m = 1$, (c) $m = 2$, (d) $m = 3$, (e) $m = 4$.

### 3.4. Autofocusing and rebound of the beams with zero and higher vorticities

The change of vorticity $m$ affects the autofocusing and defocusing (rebound) dynamics as displayed in Figs. 5 and 6. First, Figs. 5(a1-a5) demonstrate that the input with larger $m$ develops a multilayer annular shape, which is an effect of factor $r^m$ in expression (2). Next, Figs. 5(b1-b5) and 5(c) demonstrate that, as argued above, the increase of $m$ makes the radius of the ring in the tightest self-compressed state larger, and Fig. 5(c) also shows that the multilayer structure of the input naturally gives rise to a similar structure of the ring in the rebound state (actually, for $m = 3$ and 4 ). On the other hand, Figs. 5(a1) and 5(b1) show that the zero-vorticity beam has essentially smoother shapes in the input and tightest-compressed states.

Further, Fig. 6 corroborates that, for larger $m$, the rebound takes place at larger values of the radius of the shrinking ring. Note that, in the case of $m=0$, the self-autofocusing is complete in Fig. 6(a), as in that case the tightest self-compressed beam has no inner "hole". On the other hand, Fig. 6 also suggests that the focusing distance, at which the self-compression switches to the rebound, does not depend on $m$, including the case of $m=0$. Indeed, the inspection of numerical data demonstrates that this distance is completely independent of $m$, up to the accuracy of the data. This conclusion may be explained by the fact that the shrinkage of the ring-shaped input is driven by the fractional diffraction in the radial direction, which does not depend on the vorticity.

**4. The propagation of a pair of autofocusing vortex beams**

A natural extension of the analysis is to consider the evolution of a pair of the beams with unitary vorticities $m=1$, which may be considered as a result of splitting of a vortex with $m=2$. The respective input, replacing expression (2), is

$$u(r,\phi,z=0) = A_0 \text{Ai}\left(\frac{r_0-r}{bw}\right)\exp\left(d\frac{(r_0-r)}{bw}\right)\exp\left[-\frac{(r_0-r)^2}{w^2}\right]\left[\frac{(re^{i\phi}+\rho)(re^{i\phi}-\rho)}{w^2}\right]. \quad (3)$$

In this ansatz, the vortex factor $(r/w)^m e^{im\phi}$ with $m=2$ in Eq. (2) is replaced by the split one, $(re^{i\phi}+\rho)(re^{i\phi}-\rho)/w^2$, which places initial positions of pivots of the two unitary vortices at points $(x,y)=(\pm\rho,0)$. In the simulation presented below, we fixed $\rho=0.6$ mm, keeping values of other parameters the same as in Fig. 1.

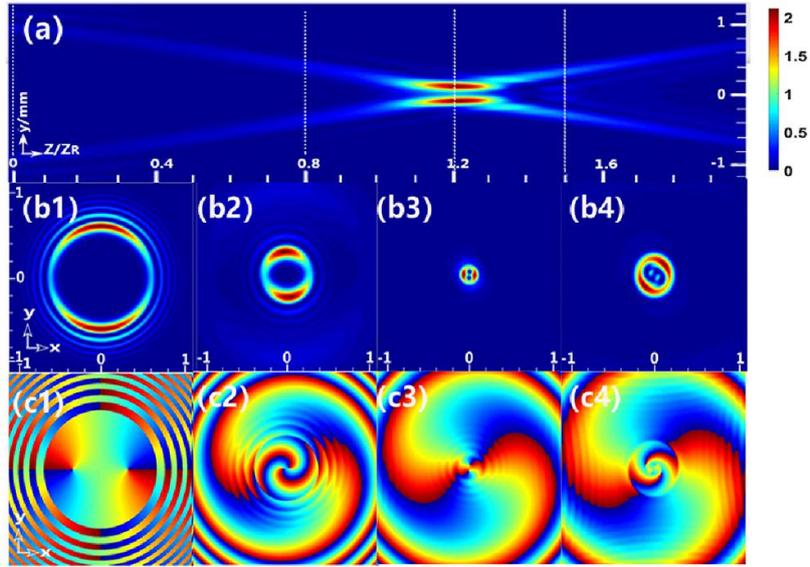

Fig. 7. (color online) The same as in Fig. 1 [and in Fig. 6(c), as concerns panel (a)], but with input (3) representing the split vortex pair.

A typical example of the autofocusing of the vortex pair is presented in Fig. 7, which explicitly displays the split structure of the double vortex in panels (b3) and (b4). In the course of autofocusing, the pair rotates in the clockwise direction. The rebound switches the autofocusing to defocusing, and simultaneously the rotation direction of the split pair switches from clockwise to counter-clockwise.

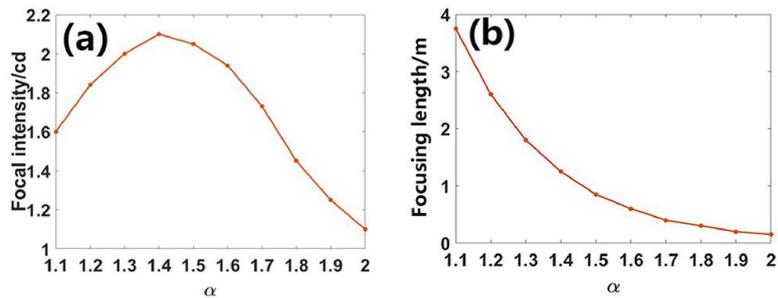

Fig. 8. (color online) (a) The focal intensity and (b) focusing length versus LI for the self-compressing split vortex pair created by input (3). The distribution parameter is $b = 0.1$.

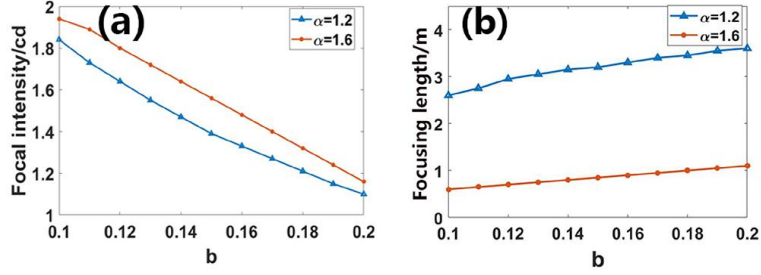

Fig. 9. (color online) (a) The focal intensity and (b) focusing length versus distribution parameter $b$ in input (3), for the autofocusing split vortex pair, at two fixed values of LI, $\alpha = 1.2$ and $\alpha = 1.6$.

To summarize the results obtained for the evolution of the vortex pair, Fig. 8 displays numerically found dependences of the focal intensity and self-focusing length on LI, cf. Fig. 2, which shows the same dependences for the unitary vortex. In particular, similar to Fig. 2(c), panel 8(a) demonstrates that the peak power in the tightest-autofocused state attains a maximum at an intermediatory value of LI, close to $\alpha = 1.4$. Further, Fig. 9 displays the dependence of the same characteristics of the autofocusing dynamics on distribution parameter $b$, cf. Figs. 3(c,d) for the unitary vortex. A notable difference is that the dependence of the self-focusing distance on $b$ is essentially weaker for the split vortex pair than for the unitary vortex beam. Indeed, the splitting makes the vortex pair broader, hence its evolution is less sensitive to details of the spatial structure of the input.

## 5. Conclusion

We have reported systematic results for the propagation dynamics of abruptly autofocusing AGV (Airy-Gaussian vortex) beams in the framework of the model of a linear optical medium based on the FSE (fractional Schrödinger equation). The initial ring-shaped beam quickly shrinks, under the action of the effective ring's elasticity, to a tightly focused structure, and then bounces back, demonstrating rapid defocusing. The beams with winding numbers from 1 to 4, studied in this work, conserve the intrinsic vorticity and the corresponding "inner hole" in the course of the autofocusing and defocusing. On the other hand, the hole disappears in the tightest self-compressed state of the zero-vorticity beam. The autofocusing dynamics is strongly affected by value of LI (Lévy index), $\alpha$, which determines the fractality of FSE. In particular, the peak intensity in the tightest-autofocused state attains a well-pronounced maximum at an intermediate value of LI, which is close to $\alpha = 1.4$. The characteristics of the autofocusing dynamics may be controlled as well by the "distribution parameter" $b$ in input (2), which measures the ratio of initial widths of the Airy and Gaussian factors in the input. Essential results are also produced for the autofocusing dynamics in vortex beams carrying higher values of the vorticity, $m$, as well as for the beams with $m = 0$. A noteworthy fact, which is explained by means of a qualitative argument, is that the smallest size of the autofocused vortex beam essentially increases with the growth of $m$. Finally, the evolution of a split vortex pair is systematically studied too, demonstrating rotation of the pair, the sign of which switches simultaneously with the rebound from the autofocusing to defocusing.

The analysis reported in this work may be extended for the FSE including a nonlinear term, which may essentially affect the resulting dynamics, as suggested by previous works which addressed effects of the nonlinearity on autofocusing of Airy waves [8, 12, 47].

**FUNDING INFORMATION**

National Natural Science Foundation of China (11775083 and 11374108); Program of Innovation and Entrepreneurship for Undergraduates; Special Funds for the Cultivation of Guangdong College Students' Scientific and Technological Innovation ("Climbing Program" Special Funds) (pdjh2020a0149). The work of B.A.M. is supported, in part, by the Israel Science Foundation through grant No. 1286/17. Science and Technology Program of Guangzhou (No. 2019050001).

**CONFLICT OF INTEREST**

We declare that we have no conflict of interest.